\documentclass[twocolumn]{aastex631}

\usepackage{xcolor}
\usepackage{lipsum}
\usepackage{placeins}
\usepackage{aas_macros}
\usepackage{amsmath,amsthm,amssymb}
\usepackage[figuresleft]{rotating}
\usepackage{afterpage}
\usepackage{capt-of}
\usepackage{rotfloat}
\usepackage[nolist,nohyperlinks]{acronym}
\usepackage{varioref}
\usepackage{hyperref} 
\hypersetup{
  colorlinks=true,   
  urlcolor=blue,     
  linkcolor=blue,
  citecolor=blue 
}

\usepackage{array}
\usepackage{vcell}
\usepackage{colortbl}
\usepackage{hhline}
\usepackage{xstring}
\usepackage{gensymb}
\usepackage{twoopt}
\usepackage{placeins}
\usepackage{changepage}
\usepackage{etoolbox}
\usepackage{cleveref}
\makeatletter
\usepackage{etoolbox}
\patchcmd\H@refstepcounter{\protected@edef}{\protected@xdef}{}{}
\makeatother

\makeatletter
\newcommand{\bibnote}[2]{\global\@namedef{#1note}{#2}}
\newcommand{\biblink}[2]{\global\@namedef{#1link}{#2}}
\makeatother

\makeatletter
  \protected\def\stonyslink{%
     \def\hyper@linkstart##1##2{}\let\hyper@linkend\@empty}
  \newcommandtwoopt{\citeads}[3][][]{%
   \href{http://ui.adsabs.harvard.edu/abs/#3/abstract}%
        {\stonyslink \citealp[#1][#2]{#3}}
   \biblink{#3}{\href{http://ui.adsabs.harvard.edu/abs/#3/abstract}{ADS}}}
 \newcommandtwoopt{\citepads}[3][][]{%
   \href{http://ui.adsabs.harvard.edu/abs/#3/abstract}%
        {\stonyslink \citep[#1][#2]{#3}}
   \biblink{#3}{\href{http://ui.adsabs.harvard.edu/abs/#3/abstract}{ADS}}}
 \newcommandtwoopt{\citetads}[3][][]{%
   \href{http://ui.adsabs.harvard.edu/abs/#3/abstract}%
        {\stonyslink \citet[#1][#2]{#3}}
  \biblink{#3}{\href{http://ui.adsabs.harvard.edu/abs/#3/abstract}{ADS}}}
 \newcommandtwoopt{\citeyearads}[3][][]{%
   \href{http://ui.adsabs.harvard.edu/abs/#3/abstract}%
        {\stonyslink \citeyear[#1][#2]{#3}}
   \biblink{#3}{\href{http://ui.adsabs.harvard.edu/abs/#3/abstract}{ADS}}}
\makeatother

\newcommand{\kw}[1]{%
\StrBehind{#1}{ (}[\result]
\StrGobbleRight{\result}{1}[\result]
\href{http://astrothesaurus.org/uat/\result}{#1}
}

\begin{document}
\title{Reliable Transmission Spectrum Extraction with a Three-Parameter Limb Darkening Law}



\author[0009-0007-1342-2498]{Rosa E. Keers}
\affiliation{Max Planck Institute for Solar System Research,\\ Justus-von-Liebig-Weg 3, 37077 Göttingen, Germany
}
\affiliation{Technische Universität Braunschweig,\\ Faculty of Electrical Engineering, Information Technology, Physics,\\ Hans-Sommer-Straße 66, 38106 Braunschweig}

\author[0000-0002-8842-5403]{Alexander I. Shapiro}
\affiliation{Max Planck Institute for Solar System Research,\\ Justus-von-Liebig-Weg 3, 37077 Göttingen, Germany
}
\affiliation{University of Graz, Institute of Physics, Universitätsplatz 5, 8010 Graz, Austria}

\author[0000-0002-6087-3271]{Nadiia M. Kostogryz}
\affiliation{Max Planck Institute for Solar System Research,\\ Justus-von-Liebig-Weg 3, 37077 Göttingen, Germany
}
\author[0000-0002-5322-2315]{Ana Glidden}
\affiliation{Department of Earth, Atmospheric and Planetary Sciences,\\
Massachusetts Institute of Technology, Cambridge, MA 02139, USA}
\affiliation{Department of Physics and Kavli Institute for Astrophysics and Space Research,\\
Massachusetts Institute of Technology, Cambridge, MA 02139, USA}

\author[0000-0002-8052-3893]{Prajwal Niraula}
\affiliation{Department of Earth, Atmospheric and Planetary Sciences,\\
Massachusetts Institute of Technology, Cambridge, MA 02139, USA}

\author[0000-0002-3627-1676]{Benjamin V. Rackham}
\affiliation{Department of Earth, Atmospheric and Planetary Sciences,\\ Massachusetts Institute of Technology, Cambridge, MA 02139, USA}

\author[0000-0002-6892-6948]{Sara Seager}
\affiliation{Department of Physics and Kavli Institute for Astrophysics and Space Research,\\
Massachusetts Institute of Technology, Cambridge, MA 02139, USA}

\affiliation{Department of Earth, Atmospheric and Planetary Sciences,\\ Massachusetts Institute of Technology, Cambridge, MA 02139, USA}

\affiliation{Department of Aeronautics and Astronautics,\\
MIT, 77 Massachusetts Avenue, Cambridge, MA 02139, USA}

\author[0000-0002-3418-8449]{Sami K. Solanki}
\affiliation{Max Planck Institute for Solar System Research,\\ Justus-von-Liebig-Weg 3, 37077 Göttingen, Germany
}

\author[0000-0001-8217-6998]{Yvonne C. Unruh}
\affiliation{ Department of Physics, Imperial College London,\\ London SW7 2AZ, UK}

\author[0009-0009-3020-3435]{Valeriy Vasilyev}
\affiliation{Max Planck Institute for Solar System Research,\\ Justus-von-Liebig-Weg 3, 37077 Göttingen, Germany
}

\author[0000-0003-2415-2191]{Julien de Wit}
\affiliation{Department of Earth, Atmospheric and Planetary Sciences,\\ Massachusetts Institute of Technology, Cambridge, MA 02139, USA}

\begin{abstract}
Stellar limb darkening must be properly accounted for to accurately determine the radii of exoplanets at various wavelengths.
The standard approach to address limb darkening involves either using laws with coefficients from modelled stellar spectra or determining the coefficients empirically during light curve fitting of the data. Here, we test how accurately three common laws --- quadratic, power, and a three-parameter law --- can reproduce stellar limb darkening at different wavelengths and across a broad range of stars. We show that using a quadratic limb darkening law, which is most frequently employed by the community, leads to wavelength-dependent offsets in retrieved transmission spectra.
For planets with high impact parameters ($b$ larger than about 0.5) \ {the amplitude of these offsets can reach 1\% of the transit depth which is some cases is comparable to and can even exceed the expected signals from the planetary atmosphere. Furthermore, the quadratic law causes an offset in the value of the impact parameter when it is determined by fitting the broadband transit light curves. } In contrast, using the Kipping--Sing three-parameter law leads to robust retrievals. We advocate the use of this law in retrievals, especially for transits with large impact parameters.

\end{abstract}


\begin{acronym}[MPC] 
\acro{JWST}{James Webb Space Telescope}
\acro{MHD}{magnetohydrodynamics}
\acro{TESS}{Transiting Exoplanet Survey Satellite}
\acro{HST}{Hubble Space Telescope}
\end{acronym}
\section{Introduction} \label{section:intro}
Planetary transit light curves are affected by the darkening of the planet's host star towards the limb, which occurs because oblique viewing angles near the limb probe higher and cooler layers of stellar photospheres \citep[see, e.g.,][p. 611]{HubenyandMihalas2014}. The limb darkening alters both the transit profile and transit depth so it must be taken into account to accurately determine planetary radii through transit photometry and atmospheric composition via transmission spectroscopy.

The community has created a number of theoretical datasets \citep[see, e.g.,][]{Claret2017, Claret2018, Maxted2018, Morello2022, Nadiia2022, Nadiia2023} that pre-tabulate limb darkening 
coefficients as a function of stellar fundamental parameters, i.e., effective temperature, metallicity, and surface gravity. The coefficients in these datasets are computed from different numerical models that either treat stellar atmospheres in 1D, such as {\tt PHOENIX} \citep{Husser2013}, {\tt MARCS} \citep{Gustafsson2008}, { \tt ATLAS9} \citep{Kurucz2005}, and {\tt MPS-ATLAS} \citep{Witzke2021}, or in 3D, such as {\tt Stagger} \citep{Stagger-IV:LD}. 

Yet, despite the complexities of the currently used stellar models, the increasing precision of transit observations reveal that more sophisticated models are still necessary \citep{Espinoza2015}. The shortcomings of the models have been made clear by recent Kepler, \ac{TESS}, and \ac{JWST} 
transit data that indicated that stars appear brighter at the limb than predicted by the models \citep{Maxted2018, Maxted2023, JWST_WASP-39b_NIRSPEC_PRISM}. In particular, the analysis of the \ac{JWST}/PRISM data by \citet{JWST_WASP-39b_NIRSPEC_PRISM} revealed that WASP-39 is 6\% brighter at the limb than models predict. 

Recently, \citet{Ludwig2023} and \citet{Nadiia2024} proposed that stellar surface magnetic fields might explain this discrepancy between observed and modelled limb darkening. This implies that 3D radiative \ac{MHD} simulations of stellar atmospheres for a broad range of stars are needed before theoretical limb darkening calculations can be reliably used. Such 3D radiative \ac{MHD} simulations are presently scarce so the community must rely on empirical determinations of the limb darkening. In other words, the planetary and limb darkening parameters are simultaneously fit to the transit light curve.

For such a fit the community adopts certain laws (with usually $2{-}4$ free parameters) to approximate the limb darkening \citep[see, e.g.,][]{Espinoza2016}. The most common law is a quadratic function of $\mu$, the cosine of the angle between the line-of-sight and the local stellar surface normal \citep[e.g.,][]{Linnell1984, Wade1985, Claret2000}. If the adopted law can accurately represent limb darkening, one can expect a reliable and correct determination of the planetary radius. However, if the limb darkening profile cannot be accurately described by the adopted law, then the fitting routine might compensate for the inaccuracy of the law by changing the planetary radius \citep{Espinoza2015}. The transit light curves are usually analysed over a broad range of wavelengths, corresponding to very different limb darkening profiles. The accuracy of the adopted law is expected to vary with wavelength \citep{Morello2017}, potentially leading to spurious changes in the retrieved planetary radius. 

In this paper, we investigate the magnitude of these features and show that, if not corrected, they can significantly affect the interpretation of the unprecedentedly precise \ac{JWST} transit data.

We describe our methodology in \Cref{section:methods}, with \Cref{section:forward} outlining our forward modelling of the transit light curves. In \Cref{section:LDlaws,section:more_LD_laws} we describe the limb darkening laws examined in this work. 
\Cref{section:retrieval} includes our light curve fitting procedure. 
We present our findings in \Cref{section:results}, with \Cref{section:central_transit} describing our results for a centrally transiting planet ($b = 0$) and \Cref{section:high_impact_transit} for a high impact parameter transit ($b = 0.7$). In \Cref{section:spectralimpact} we show the impact different limb darkening laws have on obtained planetary spectra. In  \Cref{subsect:impact} we showcase the effect of limb darkening law on impact parameter determined from fitting broadband transit light curves. We present our conclusions in \Cref{section:discussion}.

\section{Methods}\label{section:methods}
\subsection{Forward Modelling of the Transit Light Curves}\label{section:forward}
For our analysis, we utilised recently published stellar limb darkening spectra \citep{Nadiia2022, Nadiia2023} computed with the {\bf M}erged {\bf P}arallelized {\bf S}implified-ATLAS ({\tt MPS-ATLAS}) code \citep{Witzke2021} for stars with different fundamental parameters. {\tt MPS-ATLAS} employs a generalised version of the opacity distribution functions (ODF) approach \citep{HubenyandMihalas2014, cernetic_odf_2019, Anusha_odf_2021}, wherein opacities are initially computed on a high-resolution wavelength grid (with a resolving power of $R=500,000$) and subsequently rebinned to a lower resolution with $R \approx 400$.

{\tt MPS-ATLAS} incorporates the transfer of energy by convection utilizing mixing-length theory \citep{bohm-vintense1958}. It also enables convective overshooting \citep{kurucz1970}, which is important for limb darkening computations \citep[see details in][]{Witzke2021, Nadiia2022}. We took limb darkening profiles from `Set Two' of \citet{Nadiia2023}. They were computed with the chemical abundances from \citet{asplund2009} and the dependence of the mixing length parameter $\alpha$ (which characterises the efficiency of convective energy transport) on stellar fundamental parameters from \citet{viani2018}.

\begin{figure*}
    \centering
    \includegraphics[scale=0.45]{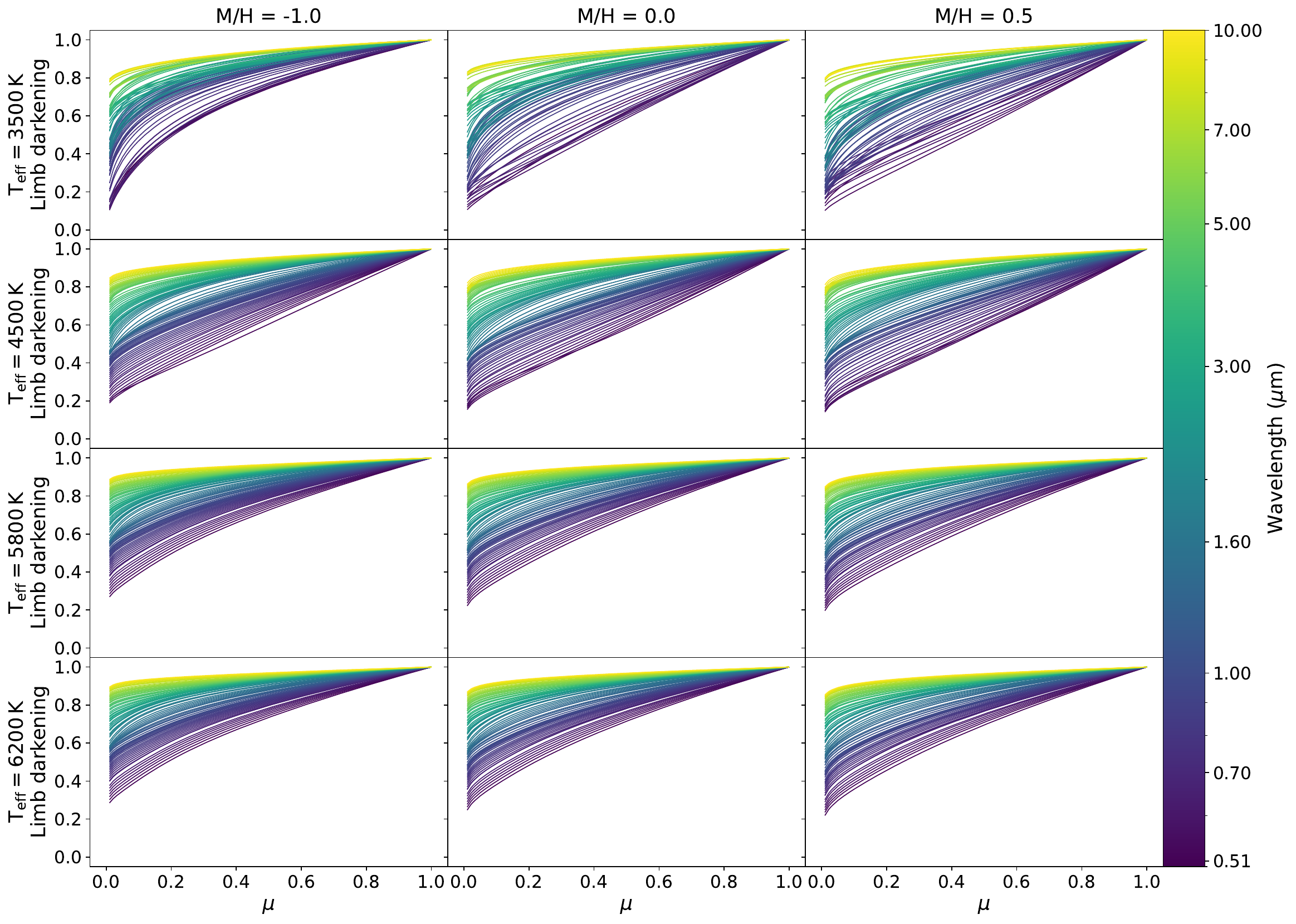}
    \caption{Stellar limb darkening curves ($I_{\lambda}(\mu)/I_{\lambda}(\mu=1))$ at different wavelengths derived from the {\tt MPS-ATLAS} spectral library \citep{Nadiia2023} for the selected stars with different fundamental parameters. The {\tt MPS-ATLAS} limb darkening has been rebinned to a spectral grid with $R\approx 40$. The $x$-axis shows $\mu$, which is the cosine of the angle of between line of sight and the stellar surface normal. The $y$-axis is the limb darkening normalised to the stellar disk center.
   The three columns display limb darkening of stars with different metallicities: $\rm{M/H}=-1.0$ (left column), $\rm{M/H}=0.0$ (middle column), and $\rm{M/H}=0.5$ (right column).
The rows illustrate limb darkening of stars with different effective temperatures, ranging from $3500$\,K to $6200$\,K (from top to bottom). The colours represent wavelengths from $0.5$\,$\mu$m to $10$\,$\mu$m, as indicated by the logarithmic colourbar. Limb darkening is wavelength dependent, occurs over the entire radius of the star, and is highly dependent on the stellar metallicity and temperature.}
    \label{fig:12_star_grid}
\end{figure*}

To encompass the diversity of stars known to host exoplanets, we have selected stars of various spectral types: F, G, K, and M, characterised by effective temperatures\,($T_{\rm eff}$) of $6200$\,K, $5800$\,K, $4500$\,K, and $3500$\,K, respectively. 
Additionally, we have included stars with different metallicities, such as metal-rich stars ($\rm{M/H} = 0.5$), metal-poor stars ($\rm{M/H} = -1.0$), and stars with solar metallicity ($\rm{M/H} = 0.0$). Limb darkening is wavelength dependent and is also affected by the stellar parameters, as shown in \Cref{fig:12_star_grid}. In general, limb darkening curves steepen at shorter wavelengths and flatten towards the infrared for all stars due to the transition from the Wien to the Rayleigh--Jeans regime \citep{Planck_paper}. Some classes of stars also exhibit more complex behaviours. In particular, for metal-rich M\,dwarfs different limb darkening curves start to cross each other (see top right panel of \Cref{fig:12_star_grid}).

We developed a simple transit model to simulate a planetary transit along a chord in front of its host star at a given impact parameter. The model includes parameters for the planet-to-star radius ratio $\frac{R_p}{R_*}$, the impact parameter $b$, the limb darkening values, and the corresponding $\mu$-values. 
We assumed a chord across the stellar disk for the trajectory of the planet since in this analysis we focus on the accuracy of the limb darkening laws without conflating it with the impact of extra orbital parameters. 

All transits were calculated for $300$\,points along a transit chord spanning $-1.4$\,$R_*$ to $1.4$\,$R_*$, with a constant planet-to-star radius ratio $\frac{R_p}{R_*} = 0.1$.
Furthermore, transits were simulated for five different impact parameters, at $b = 0.0, 0.1, 0.5, 0.7$ and $0.9$. 

The transit model broadly functions as follows: the $\mu$-values corresponding to the limb darkening values from {\tt MPS-ATLAS} are transformed to radial positions along the stellar limb. Each radius denotes the middle of a shell corresponding to a single limb darkening value. 
For each position of the planet during the transit, the model then calculates the surface of the planet which overlaps with a given limb darkening shell.
These surface areas are then combined with the limb darkening values in the shells to calculate the stellar flux blocked by the planet.\\

\subsection{Limb Darkening Laws}\label{section:LDlaws}
The main goal of this study is to assess the performance of the standard limb darkening laws, i.e. limb darkening from transit light curves, for retrieving planetary transmission spectra. For this, we proceed in the following steps. First, we synthesise the transit light curves with the forward model described in \Cref{section:forward} (i.e., using theoretical limb darkening without applying any parameterised laws). Second, we use these light curves to retrieve transmission spectra and limb darkening parameters using different limb darkening laws.

We investigated the efficacy of three limb darkening laws to correctly recover the planetary spectrum. First, we test the most commonly used law in transit spectroscopy, the quadratic law \citep{Kopal1950}:
\begin{equation}\label{eq:LD_quadratic}
    \frac{I_{\lambda}(\mu)}{I_{\lambda}(\mu=1)} = 1 - u_1 (1 - \mu) - u_2 (1 - \mu)^2.
\end{equation}
In our work, we used a version of the quadratic law where $u_{1,2}$ were re-parameterised to $q_{1,2}$ \citep[see, Eqs. 15--18 from][]{Kipping2013}.

The second law we tested was the power law \citep{Hestroffer1997}:
\begin{equation}\label{eq:LD_power2}
    \frac{I_{\lambda}(\mu)}{I_{\lambda}(\mu=1)} = 1 - c (1 - \mu^{\alpha}).
\end{equation}\\
Finally, we use a version of the \citet{Sing2009} three-parameter law:
\begin{equation}\label{eq:LD_3coeffs}
    \frac{I_{\lambda}(\mu)}{I_{\lambda}(\mu=1)} = 1 - c_2 (1 - \mu) - c_3 (1 - \mu^{\frac{3}{2}}) - c_4 (1 - \mu^2),
\end{equation}
where $c_{2,3,4}$ were re-parameterised into $\alpha_{h, r, \theta}$ according to \citet{Kipping_3coeffs}.
For clarity, we keep all parameter indices the same as in \citet{Sing2009}. Thus, the parameter indices for the three-parameter law start at two because the three-parameter law is adapted from the four parameter law \citep{Claret2000}, by removing the component associated with $c_1$ \citep{Sing2009}.
In the following, we will call the limb darkening described by \Cref{eq:LD_3coeffs} with re-parameterisation by \citet{Kipping_3coeffs} the Kipping--Sing limb darkening law.

\subsection{Different Parametrisations of the Quadratic Law}\label{section:more_LD_laws}
\ {For completeness, we also tested the original, un-reparameterised quadratic law by \citet{Kopal1950} for centrally transiting planets. This law produced results that were virtually identical (to within less than $10^{-2}$\,ppm for all wavelengths) to those produced by the \citet{Kipping2013} re-parameterisation, and as such this law was not analysed further. We construe this agreement as a confirmation that our retrieval setup (see \Cref{section:retrieval}) allows an accurate sampling throughout the entire space of retrieved parameters.}

\ {Our findings contrast with those of \citet{BiasinLDPaper2024Coulombe}, who reported that their fit planetary radius was biased by the quadratic law parameterisation used in their setup. We suggest that the difference might be the result of either their different fitting set-up (see \Cref{section:retrieval} for details on our set-up, specifically regarding the sampler) and/or their choice of unphysical limb darkening coefficients in their synthesised transit light curves.}


\subsection{Light Curve Fitting: Retrieval of Transmission Spectra and Empirical Limb Darkening}\label{section:retrieval}

We fit our light curves using the Markov chain Monte Carlo (MCMC) Ensemble sampler {\tt emcee} \citep{emcee} using the same transit model used to create the light curve (see \Cref{section:forward}). The limb darkening input to the transit model is now replaced with one of the limb darkening laws from \Cref{section:LDlaws}. We then fit for the planet-to-star ratio $\frac{R_p}{R_*}$ and the limb darkening coefficients for the selected law. The correct impact parameter is provided to the transit model, instead of being fit. This is done on the basis that orbital parameters such as impact parameter are typically retrieved from white light curves and not during spectroscopic retrievals.

For each light curve fit, the sampler was run for $5000$\,steps, with $32$\, walkers per free parameter. We set the burn-in time to three times the largest of the free parameters' autocorrelation times.

We employed uniform priors on all parameters, which are between $0$ and $1$ for nearly all parameters. The exceptions were the planet-to-star radius ratio $\frac{R_p}{R_*}$, which ranged from $10^{-4}$ to $1$, and the second parameter $\alpha$ of the power law, which ranged from $0$ to $5$.
\begin{figure*}
    \centering
    \includegraphics[width=\textwidth]{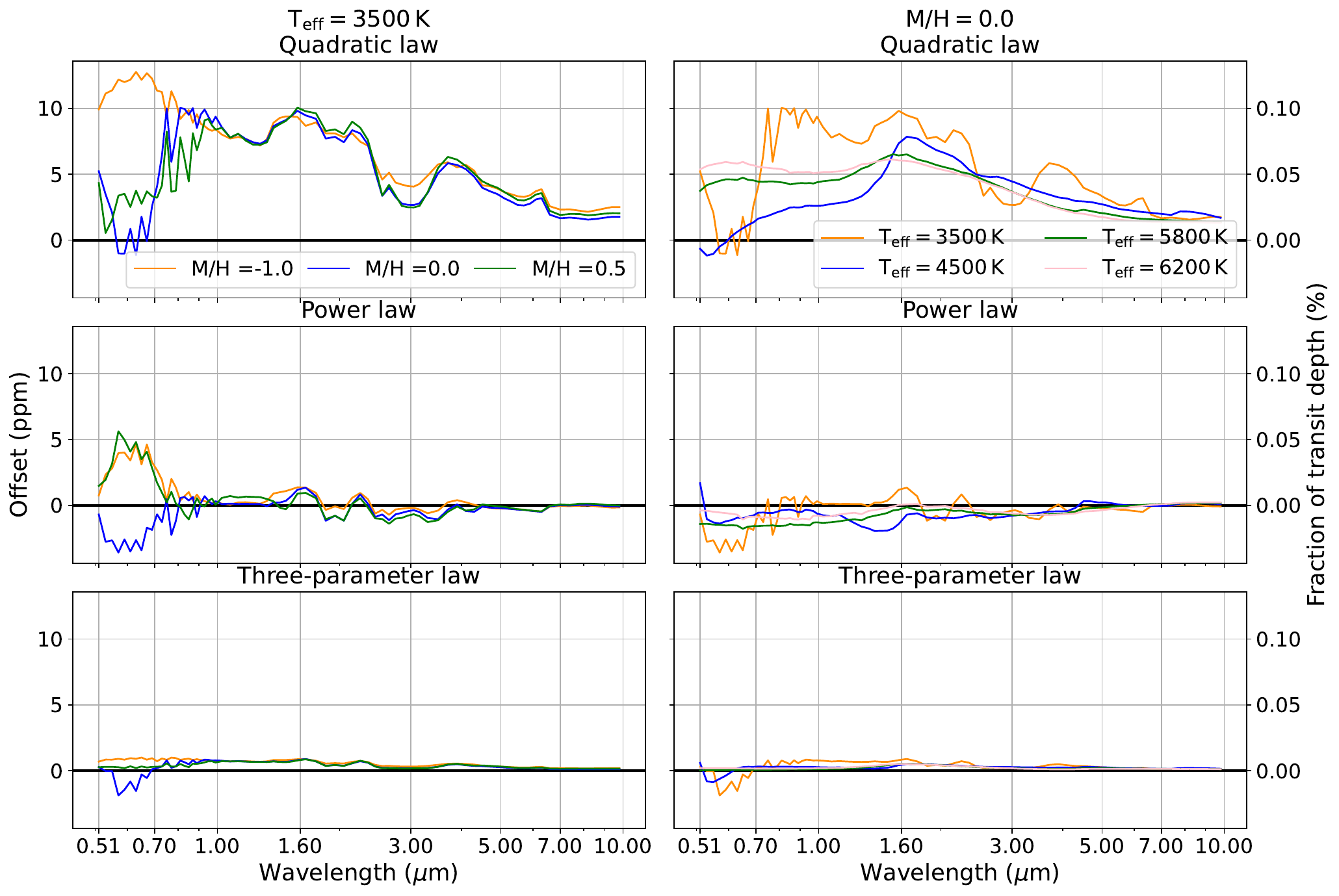}
    \caption{Centrally transiting planets ($b = 0$) wavelength-dependent offset from modelled transit depth. The left column shows offsets for stars with a constant effective temperature $T_{\rm{eff}} = 3500$\,K, and varying metallicity $\rm{M/H} = [-1.0, 0.0, 0.5]$. The figures in the right column show these offsets for stars with a constant solar metallicity $\rm{M/H} = 0.0$, and varying effective temperatures $T_{\rm{eff}} = [3500, 4500, 5800, 6200]$\,K. The thick black lines at zero offset correspond to the true transit depth. As in \Cref{fig:12_star_grid}, the wavelengths are scaled logarithmically. For both columns, each panel is for a different limb darkening law: quadratic law (top), power law (middle) and Kipping--Sing three-parameter law (bottom). The quadratic law introduces up to $13$\,ppm offsets from the modelled transit depth in the visible.
    }
    \label{fig:six_b0.0}
\end{figure*}

We performed several tests on our light curve retrieval, first by simulating several central transits with limb darkening calculated using the limb darkening laws in \cref{eq:LD_quadratic,eq:LD_power2,eq:LD_3coeffs} instead of {\tt MPS-ATLAS}. When retrievals were performed using these transits and their corresponding laws, the transit depth was accurately constrained down to $\approx 0.1$\,ppm.

Second, we assessed our model's consistency by repeatedly retrieving spectra from the {\tt MPS-ATLAS} generated transit light curves. If the retrievals displayed smooth and reasonable differences within the range of 10 to 100 ppm, we selected the last of our retrievals to avoid biasing our results. These smooth variations likely stem from the limitations of synthesizing transit light curves with limb darkening laws, which may not perfectly match theoretical computations. In real exoplanet transit spectrum retrievals, it would be impossible to pinpoint the `true' result. However, we also observed extreme jumps in transit depth at a single wavelength, ranging from 1000 to 100,000 ppm, which coincided with significant changes in limb darkening coefficients. These extreme, nonphysical jumps were considered suspicious and matched notable changes in the limb darkening parameters, prompting us to repeat these retrievals until the extreme values were eliminated.

Finally, we also repeated the central transit retrievals ($b = 0$), using the quadratic and power laws, for a smaller planet with $\frac{R_p}{R_*} = 0.01$. The spectra this produced matched those at $\frac{R_p}{R_*} = 0.1$ in shape, but as expected only had $\frac{1}{100}$ of the amplitude of those of the larger planet.

In the following section, we present the results of our retrievals in terms of the deviation from the true, flat transit spectrum.
We define the offset from the true transit depth as 
$(\frac{R_p}{R_*})^2_{\rm{Fit}} - (\frac{R_p}{R_*})^2_{\rm{True}}$. 
We also translate this to a fraction of the (true) transit depth, as $\frac{(\frac{R_p}{R_*})^2_{\rm{Fit}} - (\frac{R_p}{R_*})^2_{\rm{True}}}{(\frac{R_p}{R_*})^2_{\rm{True}}}$.
Thus, a positive offset indicates that the planet seems larger than it truly is, while a negative offset indicates a smaller $\frac{R_p}{R_*}$ than reality.

\section{Results}\label{section:results}

Previous studies have already examined the effects arising from  approximating limb darkening by different laws on the analysis of the photometric transit light curves. In particular, \citet{Morello2017} has shown that the use of the quadratic law as well as other two-parameter laws
leads to significant biases in the retrieved planetary radii.

In this section, we, for the first time, investigate  the impact of different limb darkening laws on planetary spectra across a broad wavelength range. \ {We find that this impact has two main components. First, the inability of a chosen law to precisely describe the limb darkening at a given wavelength is compensated by offsetting the planetary radius (with respect to the true value) during the fitting of the spectral transit light curves. This effect introduces wavelength-dependent offsets in the transmission spectra and is described in \Crefrange{section:central_transit}{section:spectralimpact}. Second, the usage of the limb darkening laws leads to an offset in the value of the transit impact parameter when it is defined from the white (or more precisely broadband) light curves as it is routinely done in the literature. The error in the impact parameter then propagates into
the transmission spectrum since the fitting routine tries to compensate for the wrong impact parameter by offsetting the planetary radius. We discuss this effect in \Cref{subsect:impact}.}



\subsection{Centrally Transiting Planets: Well Described by Kipping--Sing Three-Parameter Law}\label{section:central_transit}

The widely used quadratic limb darkening law impacts transit depth by up to $13$\,ppm (i.e. 0.12\% of the fractional transit depth) for central transits ($b=0$). Furthermore, the offsets induced by the quadratic limb darkening law are very rich in spectral features for M\,dwarfs due to strong molecular bands in their spectra. On the contrary, the Kipping-Sing three-parameter law accurately describes limb darkening to within a $2$\,ppm offset from the transit depth (for $b=0$). 

This result is showcased in \Cref{fig:six_b0.0}. The left panels of \Cref{fig:six_b0.0} show the transit depth offsets for a central transit across three stars with constant stellar effective temperature ($T_{\rm{eff}} = 3500$\,K) and variable metallicity ($\rm{M/H} = -1.0$, $0.0$ or $0.5$). 
The right panels of \Cref{fig:six_b0.0} show the same central transit, but now for stars with a constant solar metallicity ($\rm{M/H} = 0.0$), and varying effective temperature ($T_{\rm{eff}} = 3500$\,K, $4500$\,K, $5800$\,K or $6200$\,K).
In \Cref{fig:six_b0.0}, the quadratic law performs the worst of the three laws. The power law functions notably better, especially at longer wavelengths beyond $\lambda \approx 1.0$\,$\mu$m.
Note that \Cref{fig:six_b0.0} shows that the offset has a strong dependency on metallicity at sub-infrared wavelengths.
Neither of the two-parameter limb darkening laws is able to reach under a $5$\,ppm absolute offset in the optical spectral range.
Instead, the corresponding offsets display a strong wavelength dependence, which may prove difficult to disentangle from planetary atmospheric features. 

By contrast, the absolute offset resulting from the three-parameter law remains below $2$\,ppm across the entire wavelength range. This implies that three-parameter Kipping--Sing law is capable of accurately representing limb darkening.

\subsection{High Impact Parameter Transits: Only the Kipping--Sing Three-Parameter Law Leads to Accurate Transmission Spectra}\label{section:high_impact_transit}

For high impact parameter transits, the gain in accuracy using the Kipping--Sing three-parameter law compared with the quadratic law is even more marked, with wavelength dependent features for the former around $10 {-} 20$\,ppm while for the latter they are around $100 {-} 130$\,ppm.  

\begin{figure*}
    \centering
    \includegraphics[width=1.0\textwidth]{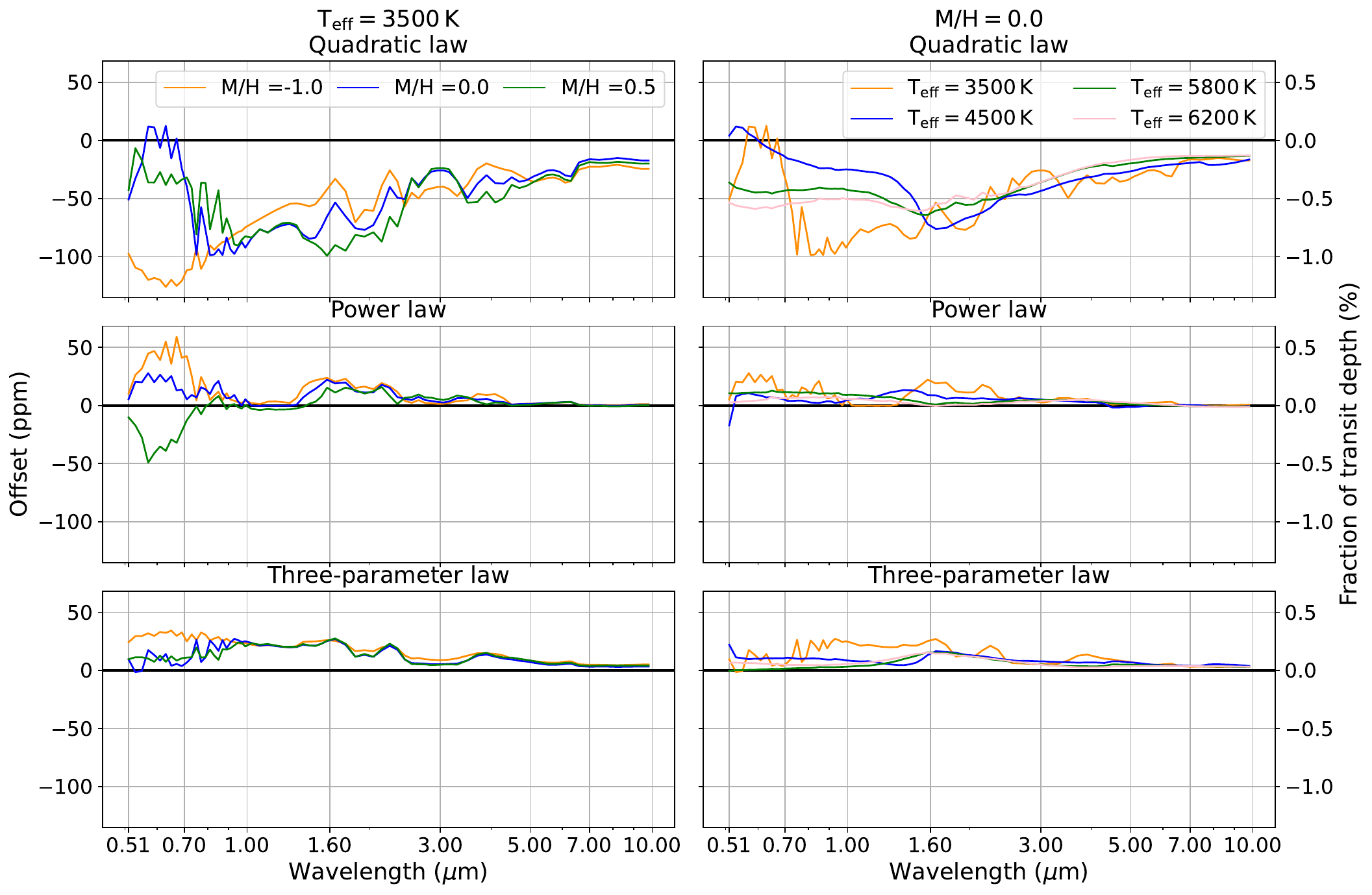}
    \caption{High impact parameter ($b = 0.7$) wavelength dependent offset from modelled transit depth. See \Cref{fig:six_b0.0}, for a detailed description of the figure lay-out. Note the increased offset, up to $130$\,ppm for the quadratic law, compared with \Cref{fig:six_b0.0}.}
    \label{fig:six_b0.7}
\end{figure*}

\Cref{fig:six_b0.7} shows the transit depth offsets for a transit at impact parameter $b=0.7$. 
Here we chose to showcase this impact parameter because it corresponds to the largest offset (notably even larger than for the grazing case of $b=0.9$, see \Cref{fig:six_b0.9} in Appendix\,\ref{app:b0.9}) when the quadratic law is used for the transit light curve fitting.  
The offset for the quadratic law reaches approximately $130$\,ppm (i.e., around $1.3$\% of transit depth and approximately $0.6$\% of planetary radius, see \Cref{fig:six_b0.7}, upper left panel). 

By contrast, the Kipping--Sing three-parameter law performed significantly better at all wavelengths.
In the bottom left panel of \Cref{fig:six_b0.7}, the three offset spectra corresponding to the metallicity values $\rm{M/H} = -1.0$, $0.0$ and $0.5$ have median absolute values of $20.2$, $10.6$ and $11.3$\,ppm respectively. This compares favourably to the median absolute values of $52.1$, $38.0$ and $40.9$\,ppm for the three respective metallicity offset spectra retrieved using the quadratic law.

\subsection{Impact of Limb Darkening Law on Transmission Spectra of Super-Earth and Warm-Jupiter}\label{section:spectralimpact}

In this section, we show that poorly chosen limb darkening laws can have an impact on transmission spectra of both super-Earth and Jupiter-sized planets.

To contextualise the signals induced by the limb darkening, we compare them to representative planetary atmospheric spectra of a super-Earth and warm-Jupiter, presented in \citet{Niraula2022}. The planet-to-star radius ratios adopted by \citet{Niraula2022} for these spectra ($\frac{R_p}{R_*}$) are $\approx 0.1$ and $\approx 0.187$, respectively. In order to compare the warm-Jupiter spectrum to the offsets due to the limb darkening approximation, we re-scaled the offsets,  from the $\frac{R_p}{R_*} = 0.1$ value used for everything in our study, to corresponding $\frac{R_p}{R_*}$ values for the simulated planets.

Both planets were simulated with M\,dwarf hosts, with respective stellar radii of $R_* = 0.1R_{\odot}$ and $R_* = 0.55R_{\odot}$, roughly corresponding to a very late and an early M\,dwarf \citep{Pecaut2013,Mamajek2022}. The limb darkening 
for both stars have been calculated using a model corresponding to effective temperature $T_{\rm{eff}} = 3500$\,K and solar metallicity ($\rm{M/H} = 0.0$). 
The {\tt MPS-ATLAS} code currently cannot be used to simulate stars with an effective temperature below $3500$\,K \citep{Witzke2021}. Furthermore, there is evidence that the mixing length approximation used in 1D models starts to fail for cool M\,stars \citep{beeck_2013, nadiiamixlength}, and more comprehensive 3D radiative (M)HD  simulations are needed to obtain their limb darkening. Thus, in the absence of a better solution, we used a $3500$\,K stellar model to represent the limb darkening on both the early and very late M\,dwarf. This implies that the offsets shown against the super-Earth spectrum were most likely underestimated. This is because the spectra of later M\,dwarfs contain more molecular features, leading to further features in the offset spectra. 

\begin{figure}
    \centering
    \includegraphics[scale=0.4]{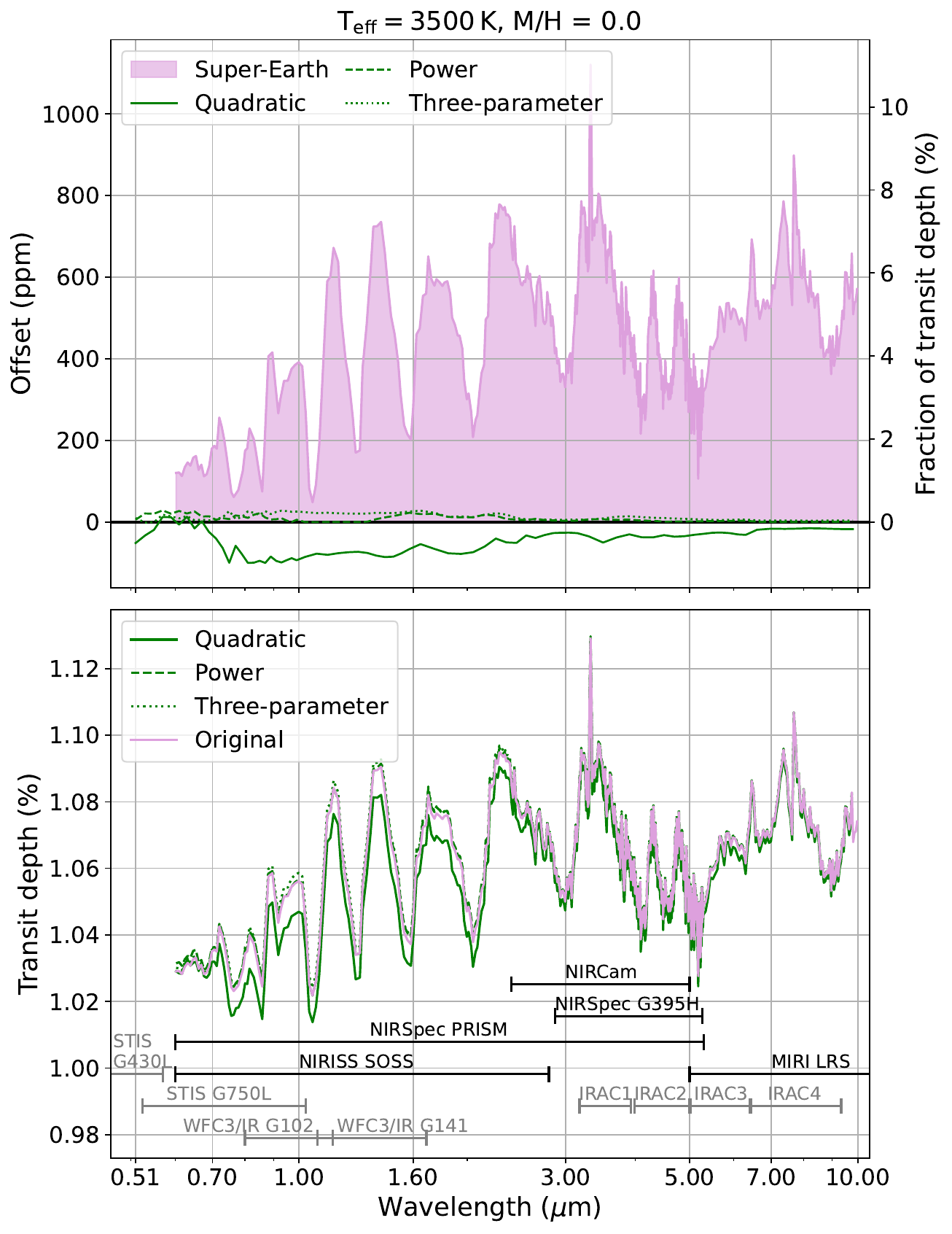}
    \caption{Top: Offsets from true transit depth induced by different limb darkening laws compared to a simulated spectrum of a super Earth from \citet{Niraula2022}.\\
    Bottom: The same spectrum of a super Earth from \citet{Niraula2022}, compared to the spectrum which would be retrieved using different limb darkening laws for the case of a $b = 0.7$ transit (see text for more details). \ac{JWST} instrument coverage is also included in black, alongside \ac{HST} and Spitzer instruments in grey. The quadratic limb darkening law results in a downward slope towards shorter wavelengths.}
    \label{fig:superearthplot}
\end{figure}

\begin{figure}
    \centering    
    \includegraphics[scale=0.4]{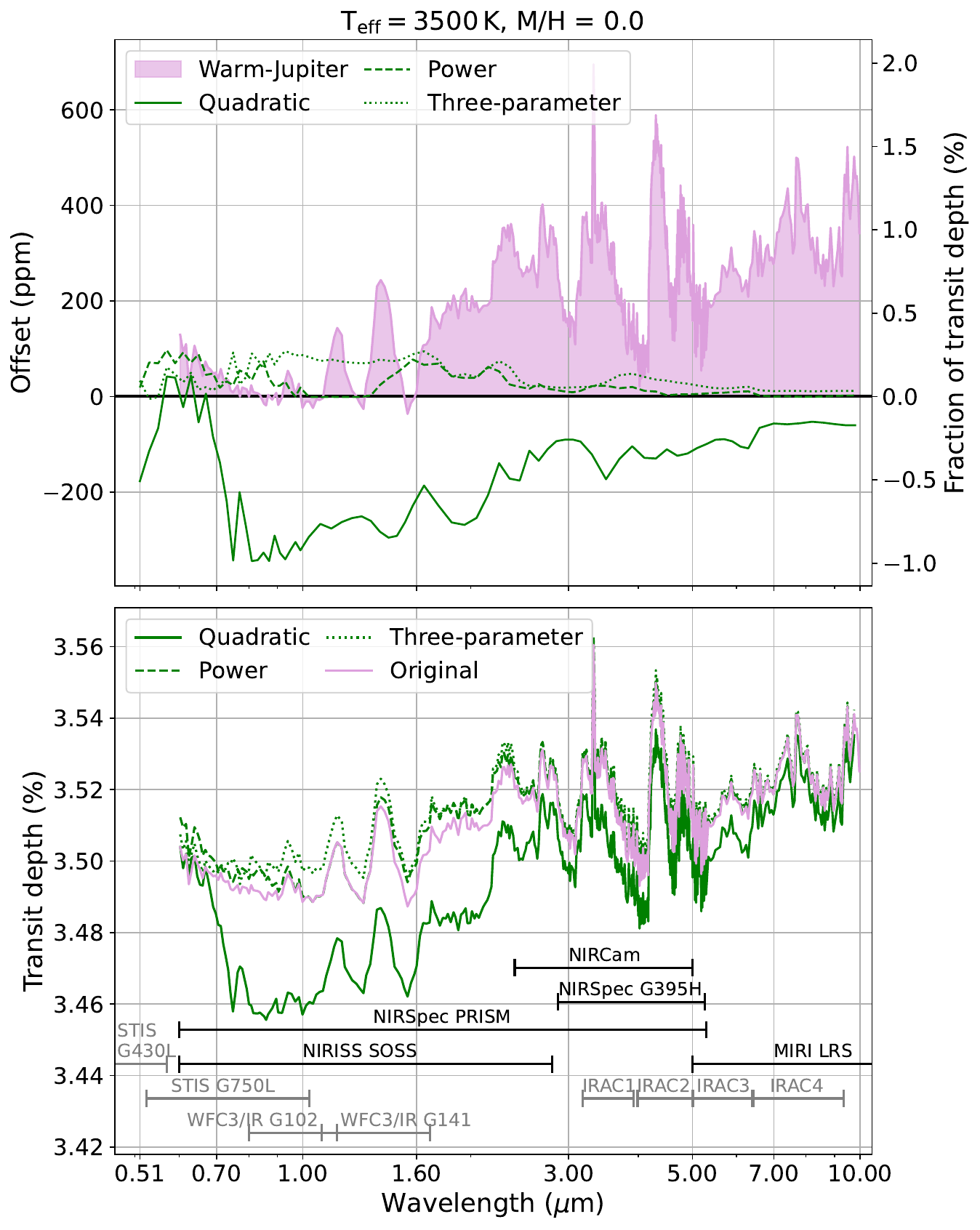}
    \caption{The same as \Cref{fig:superearthplot} but for a simulated spectrum of a warm Jupiter, from \citet{Niraula2022}. The quadratic limb darkening law leads to a downwards slope in transit depth towards shorter wavelengths and a sharp increase at $\lambda \approx 0.7$\,$\mu$m.}
    \label{fig:warmjupiter}
\end{figure}

\Cref{fig:superearthplot,fig:warmjupiter} show the effect the limb darkening law used in the retrieval has on deduced planetary spectra for the $b=0.7$ case. In the top panels, we overplot the offsets obtained with different limb darkening laws with the simulated planetary spectra. The planetary spectra have been shifted vertically, by the transit depth based on the planetary radius listed in \citet{Niraula2022} to display them analogously to the offsets from our retrievals. The bottom panels present the comparison of the planetary spectra with those retrieved with different laws (for $b = 0.7$).

The limb darkening induced offsets affect the spectra across all wavelengths, but particularly in the visible. This effect is especially extreme for the warm-Jupiter. Here a large decrease in transit depth can be seen between $\lambda \approx 0.7$\,$\mu$m and $\lambda \approx 2.5$\,$\mu$m.
While one might expect such a feature to have been noted in previous transmission spectroscopy studies, this could very well have been missed due to a combination of instrument offsets and sampling biases, as well as the rarity of large planets around cool stars.

While an in-depth investigation into all previous observations of similar warm and hot-Jupiters around cool stars goes beyond the scope of this work, we do take further note of some factors that could have led to this effect being missed.
An ideal candidate for showing the limb darkening induced effects would have both a large impact parameter, orbit a cool star, and have a large planet-to-star radius ratio. Such targets are rare, though large planets around cool stars are of great interest to the exoplanet community \citep{GEMS}. 
All in all, the ideal systems for identifying this effect, have not (yet) been observed with transit spectroscopy, e.g. TOI-1227b \citep{TESS_TOI-1227b}, TOI-3235b \citep{TESS_TOI-3235b}, TOI-5344b \citep{TESS_TOI-5344b}.
Other warm and hot-Jupiters around cool stars, which have been observed at short wavelengths, such as WASP-80b \citep{HST_WASP-80b}, have low enough impact parameters and warm enough host stars, so that any limb darkening induced slope could have 
been attributed to Rayleigh scattering.
Such degeneracies between stellar and planetary parameters warrant further investigation, but are beyond the scope of this work.

Other warm and hot-Jupiters around stars with earlier stellar types were observed in the right wavelength range(s) to catch these effects. However, the limb darkening effects are diminished for early spectral types.

\subsection{Offset in impact parameter}\label{subsect:impact}
\begin{figure}[!htp]
    \centering
    \includegraphics[scale=0.5]{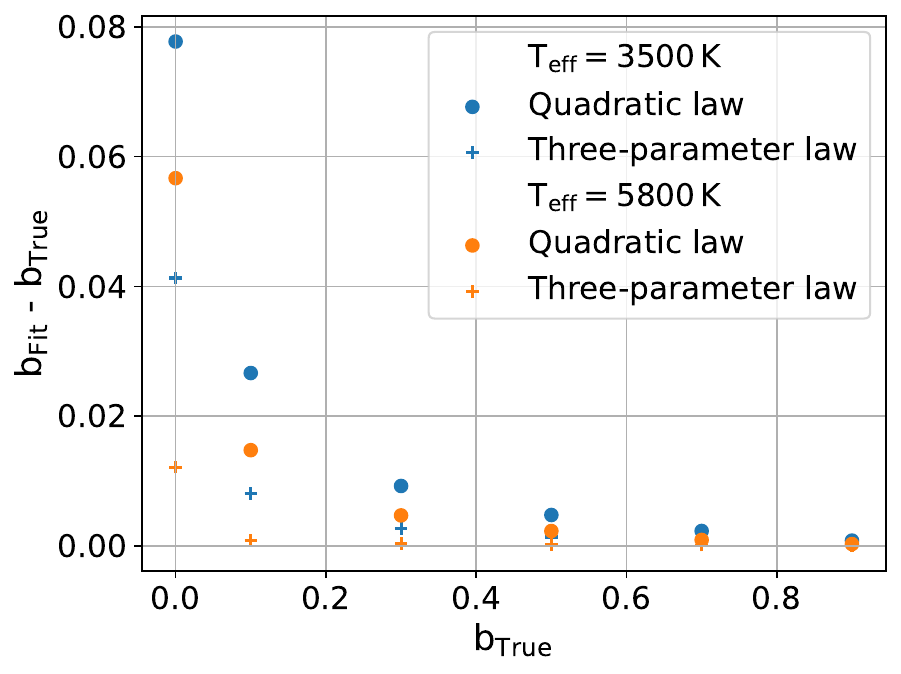}
    \caption{The difference between the true impact parameter values and those obtained from fitting light curves ($b_{\rm{Fit}}$)  versus true impact parameters ($b_{\rm{True}}$). For both limb darkening laws used to fit these impact parameters, the difference between the true and fit impact parameter values peaks for central transits.}
    \label{fig:bvsdelb}
\end{figure}

\ {In the analysis performed in \Crefrange{section:central_transit}{section:spectralimpact} we assumed that the true value of the impact parameter is known. In reality, it should be calculated from the transit light curves to be analysed and/or from complementary observations. For example, \citet{Morello2017} kept the impact parameter free during their broadband light curve fits and, as a result they reported substantially larger offsets introduced by limb darkening laws than we showcased in \Cref{fig:six_b0.0,fig:six_b0.7}. To investigate  the influence of an unknown impact parameter on the transmission spectra we will now follow a method reminiscent of that which is used during analysis of transit observations.}

\ {First, we employed our forward model to calculate a transit light curve as it would be observed in the TESS passband \citep{TESS_instrument}. Second, we then used this light curve to determine the impact parameter, planetary radius, and limb darkening coefficients. Our calculations indicate that the usage of the quadratic law during the fitting procedure leads to significant offsets in the impact parameter (see \Cref{fig:bvsdelb}). The offset is especially large for transits with low impact parameters and its amplitude also increases for cooler stars. The usage of the three-parameter law leads to a more accurate determination of the impact parameter, but still results in noticeable offsets for near-central transits.}

\ {Finally, we utilize this value of the impact parameter (in contrast to using the same value as in the forward modelling as we did previously) when fitting for the transmission spectrum. This leads to a significant increase of the magnitude of the offset transit spectrum, although it does not directly amplify the spectral features.
In the Appendix \Cref{fig:impactTS} we illustrate this finding with an example of a transit with an impact parameter of $b=0.1$. Here the power law very slightly outperforms the Kipping--Sing law, although all follow the same general trend.}



\section{Summary \& Conclusions}\label{section:discussion}
We find that spectroscopic light curve retrievals using the quadratic limb darkening law are liable to introduce (detectable) biases (of up to $130$\,ppm for $\frac{R_p}{R_*}=0.1$) in the retrieved transit depth. These biases can introduce rogue stellar spectral features in the retrieved exoplanet transit spectra. 
As such, we do not recommend the use of the quadratic law in exoplanet light curve retrievals. Instead, we find that the Kipping--Sing three-parameter law is able to more accurately describe limb darkening for nearly all cases.

The impact of choosing a limb darkening law is most noticeable for high impact parameter targets, where the Kipping--Sing Three-parameter law is able to stay below at least half the offset amplitude compared to that of the quadratic law.

When processing transits observed at long wavelengths, such as those from \ac{JWST}'s MIRI\,LRS\,($5 - 12$\,$\mu$m) \citep{MIRI_LRS, MIRI_overview}, use of the power law still produces acceptable results with very small offsets. However, we see no reason to use it going forward and prefer the Kipping--Sing three-parameter law for all limb darkening corrections. At shorter wavelengths ($\lambda \leq 5$\,$\mu$m) and high impact parameters, the retrievals using the power law also become severely biased.
For such cases, if not for all, the three-parameter law should be used.\\

Previous work, by \citet{Muller2013} using Kepler transits had indicated that empirically retrieved limb darkening could not accurately retrieve high impact parameter transits, instead recommending the use of theoretical limb darkening coefficients. This, among other factors, has led to the widespread use of theoretical limb darkening coefficients in transit retrievals, despite the numerous previously noted problems with this approach \citep{Southworth2008,Csizmadia2013, Espinoza2015, Morello2017}. These problems are due to the biases introduced by imperfections in the stellar models used, the interpolations on those model grids, as well as inaccuracies in the measured stellar parameters.

Our work shows that the three-parameter limb darkening law leads to accurate determination of transmission spectra even at high impact parameters.  Thus, it can be used for the light curve analysis before more accurate models of stellar atmospheres and limb darkening become available. Furthermore, we found that potential degeneracies in the limb darkening coefficients did not translate into offsets in planetary radius. 
Our findings agree with those of \citet{Morello2017}, who obtained accurate light curve fitting using the four-parameter law \citep{Claret2000}, provided that prior information on the orbital parameters could be used to solve some of the degeneracies.

\ {We also found that the quadratic law can lead to an inaccurate determination of the impact parameter from broadband transit light curves. These inaccuracies, in turn, can propagate to the transmission spectra \citep[see, e.g.][]{Spitzer_HD189733b_IRAC+RV} even if fitting with a more accurate limb darkening law. Thus, we emphasise that special care should be taken when determining the impact parameter (and orbital parameters in general). We advocate to use the Kipping--Sing three-parameter law and preferably obtain further constraints \citep[e.g., coming from the radial velocity measurements, see][]{Spitzer_HD189733b_IRAC+RV}.}




In the near future we plan to expand this work by using the 3D radiative MHD code MURaM \citep{Voegler2005}, instead of the 1D {\tt MPS-Atlas} code used here. The MURaM code has been demonstrated to reproduce solar observables with very high accuracy \citep{Witzke2024} and can compute valid spectra and limb darkening for very cool stars.
As such, this expansion will allow us to analyse limb darkening retrievals for transits across stars as late as M8V, such as TRAPPIST-1 \citep{TRAPPIST-1_SpType}, and also account for the effects of magnetic fields \citep{Nadiia2024}. We also plan to investigate the effects of the instrumental noise on the transit light curve analysis. 
This will furthermore involve the examination of the various methodologies employed during transit fitting, and how those influence the overall fit, parameter correlations, and sampling efficiency.

Furthermore, we plan to investigate how changing the limb darkening law affects light curve fitting of real transit data, and thus the transit spectrum, by re-processing and re-analysing previous \ac{JWST} and \ac{HST} transit spectroscopy observations. This will include several targets of high interest to the community, with large data sets available, (e.g. WASP-39b, HD\,189733\,Ab). This will also cover targets that would be ideal for investigating strong limb darkening effects, which do not have published spectroscopic transits, but currently have transit observations scheduled \citep[e.g., TOI-3235b, see][]{JWSTProp_TOI-3235b_NIRSpecPRISM}.

\ {Acknowledgments.} We thank James Thorne for the useful discussions and insightful comments. We acknowledge support form the European Research Council (ERC) under the European Union’s Horizon 2020 research and innovation program (grant no. 101118581) and the National Aeronautics and Space Administration under Agreement No.\ 80NSSC21K0593 for the program ``Alien Earths''. 
The results reported herein benefited from collaborations and/or information exchange within NASA’s Nexus for Exoplanet System Science (NExSS) research coordination network sponsored by NASA’s Science Mission Directorate.
\software{NumPy \citep{harris2020array},
Matplotlib \citep{Hunter2007},
emcee \citep{emcee}, 
SciPy \citep{SciPy-NMeth2020}
}
\clearpage
\FloatBarrier
\appendix
\FloatBarrier
\section{Highest Impact Parameter}\label{app:b0.9}
\FloatBarrier

\begin{figure*}[!htp]
    \centering
    \includegraphics[scale=0.5]{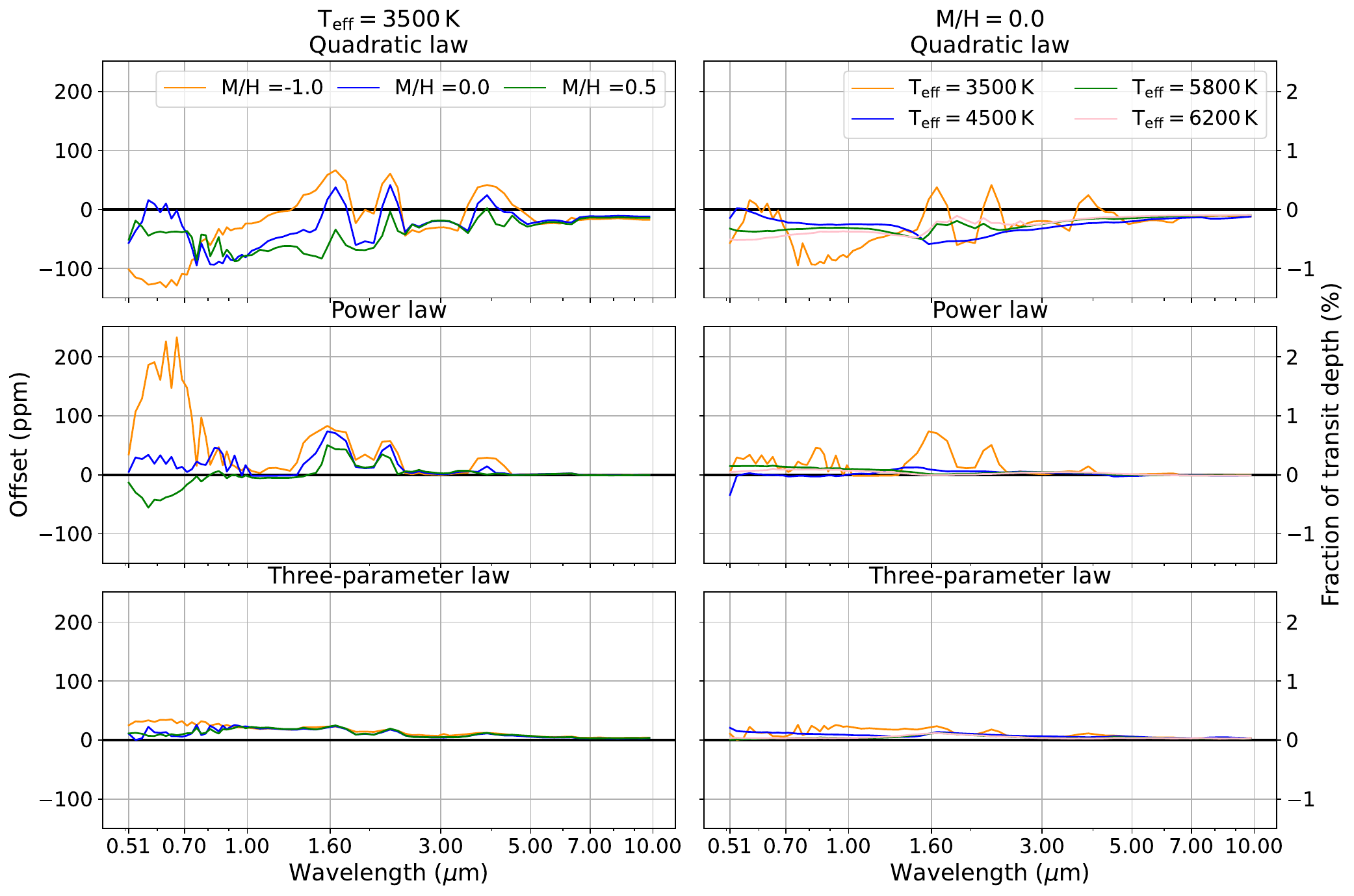}
    \caption{Offset from true transit depth over wavelength, for planets transiting at the highest impact parameter $b = 0.9$. See \Cref{fig:six_b0.0}, for a detailed description of the figure lay-out.}
    \label{fig:six_b0.9}
\end{figure*}
\newpage
\FloatBarrier
\section{Effects of impact parameter}

\begin{sidewaysfigure}[H]
    \centering
    \includegraphics[scale=0.52]{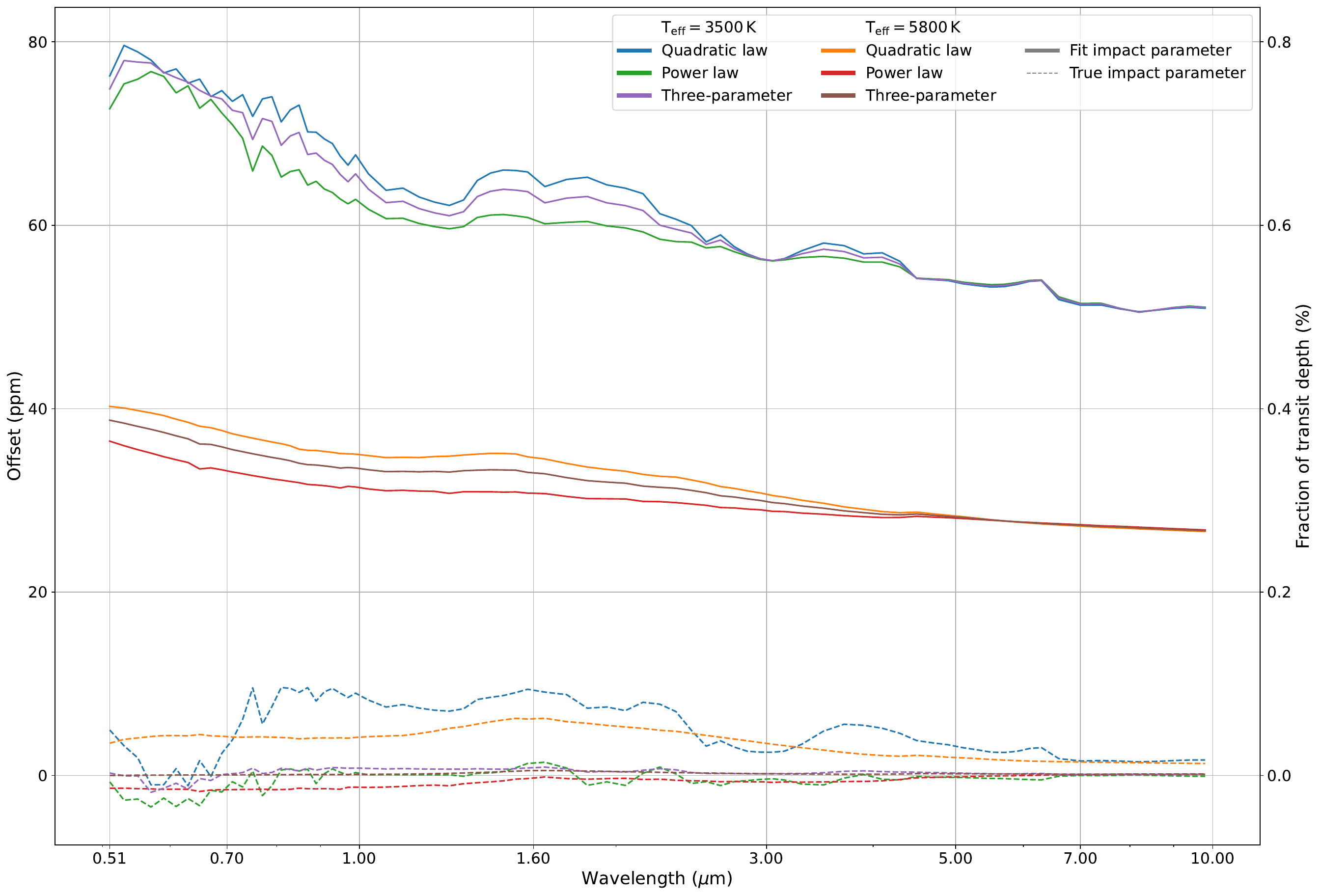}
    \caption{Offsets from true transit depth, for a planet transiting at $b = 0.1$, for both true impact parameters and those obtained from light curve fitting.}
    \label{fig:impactTS}
\end{sidewaysfigure}

\FloatBarrier
\section{Offset Spectra for full Grid}\label{section:full_grid}

\begin{sidewaysfigure}[H]
    \centering
    \includegraphics[scale=0.53]{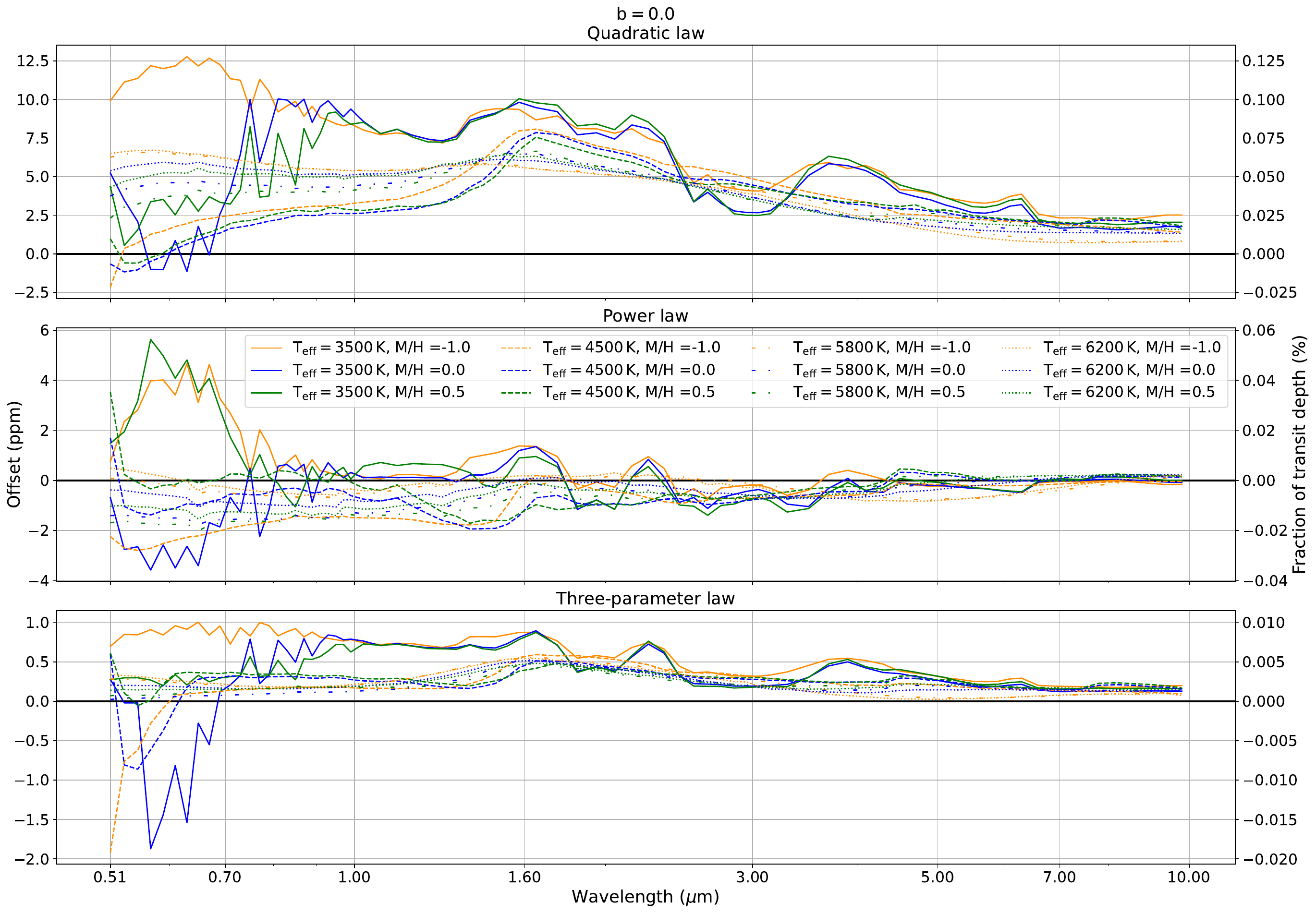}
    \caption{Offsets from true transit depth, for a centrally transiting planet, for all twelve model stars.}
    \label{fig:Teffvar_mhvar_b0.0}
\end{sidewaysfigure}

\begin{sidewaysfigure}
    \centering
    \includegraphics[scale=0.53]{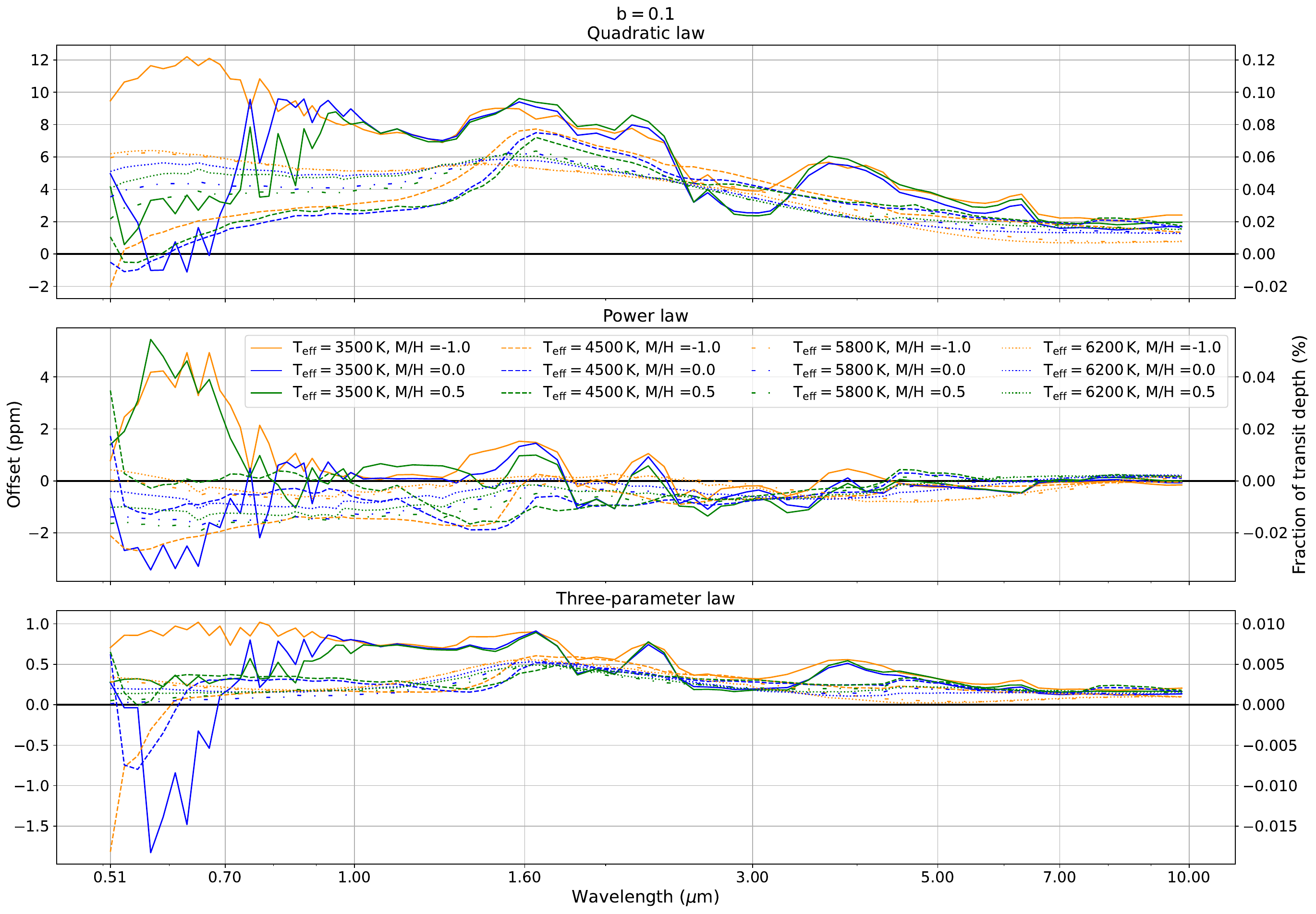}
    \caption{Offsets from true transit depth, for a planet transiting at $b = 0.1$, for all twelve stars.}
    \label{fig:Teffvar_mhvar_b0.1}
\end{sidewaysfigure}

\begin{sidewaysfigure}
    \centering
    \includegraphics[scale=0.53]{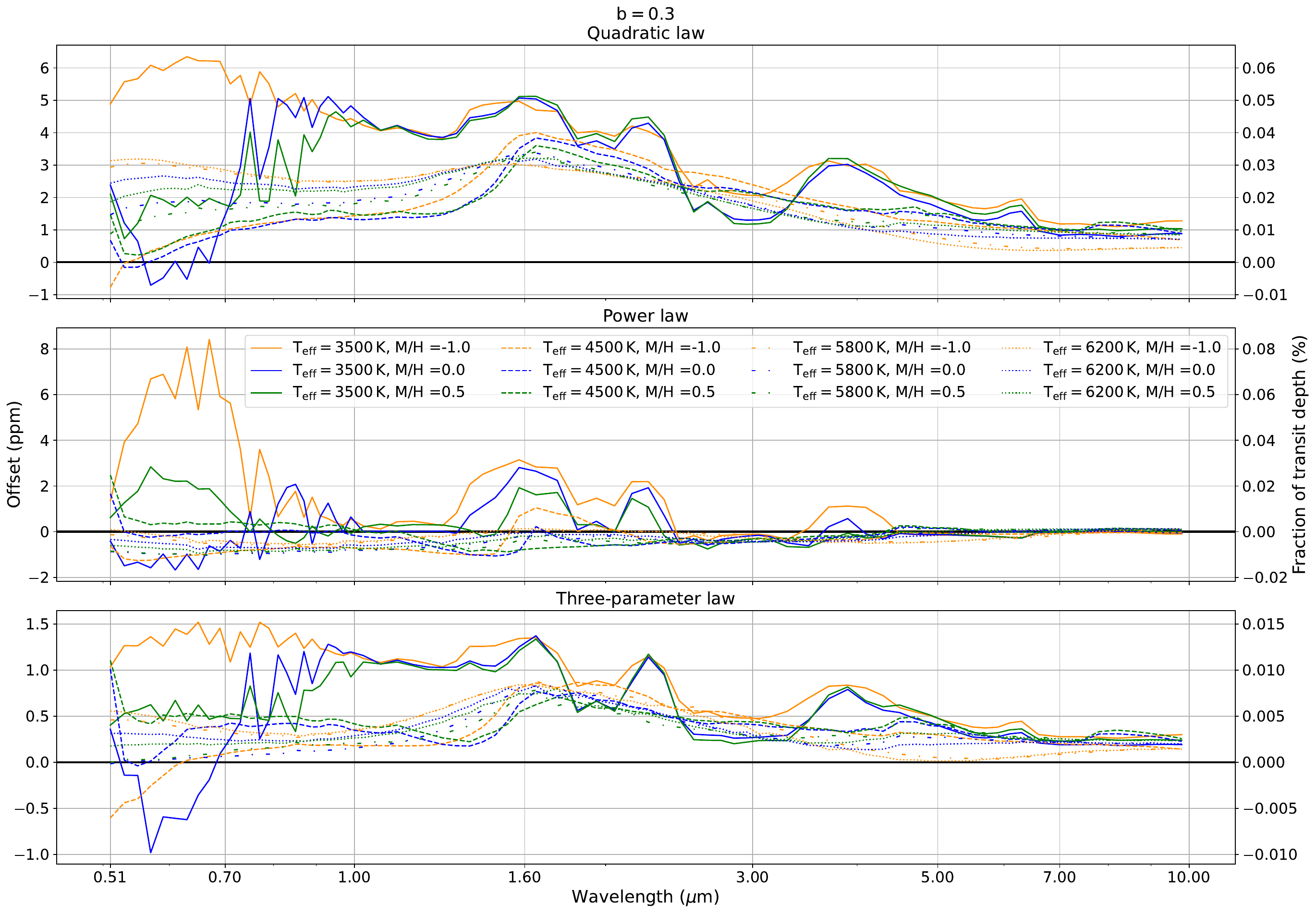}
    \caption{Offsets from true transit depth, for a planet transiting at $b = 0.3$, for all twelve stars.}
    \label{fig:Teffvar_mhvar_b0.3}
\end{sidewaysfigure}

\begin{sidewaysfigure}
    \centering
    \includegraphics[scale=0.53]{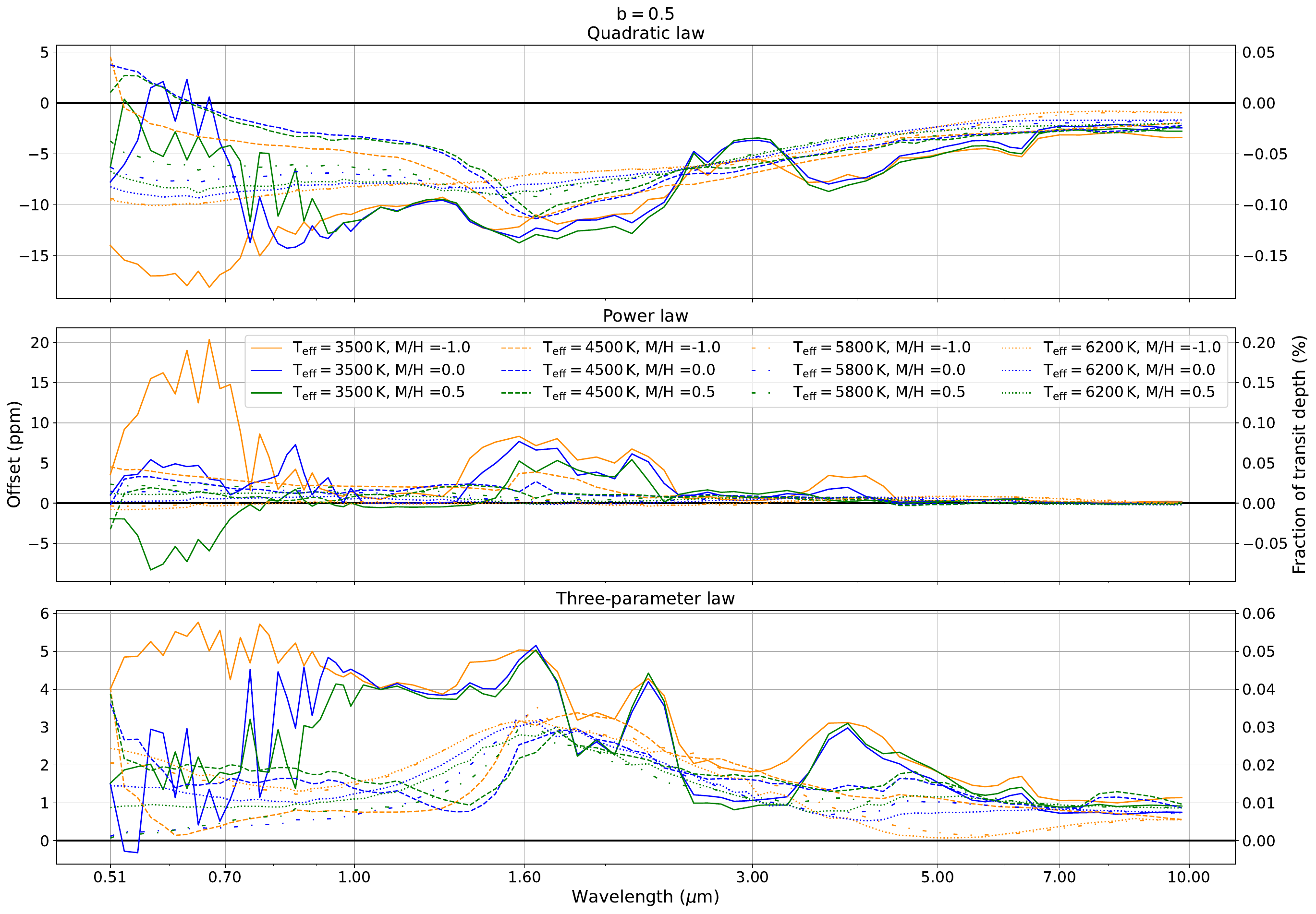}
    \caption{Offsets from true transit depth, for a planet transiting at $b = 0.5$, for all twelve stars.}
    \label{fig:Teffvar_mhvar_b0.5}
\end{sidewaysfigure}

\begin{sidewaysfigure}
    \centering
    \includegraphics[scale=0.53]{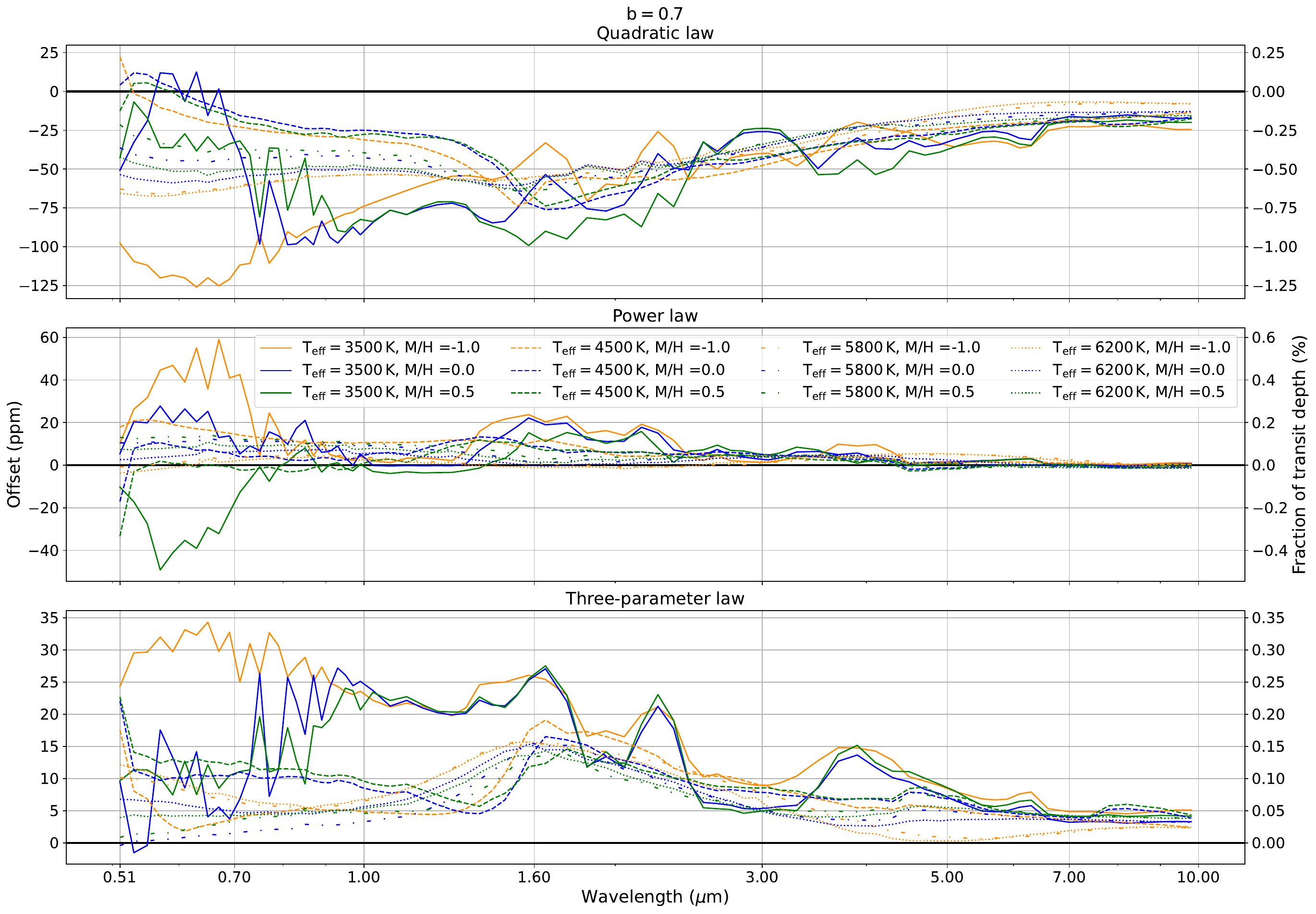}
    \caption{Offsets from true transit depth, for a planet transiting at $b = 0.7$, for all twelve stars.}
    \label{fig:Teffvar_mhvar_b0.7}
\end{sidewaysfigure}

\begin{sidewaysfigure}
    \centering
    \includegraphics[scale=0.53]{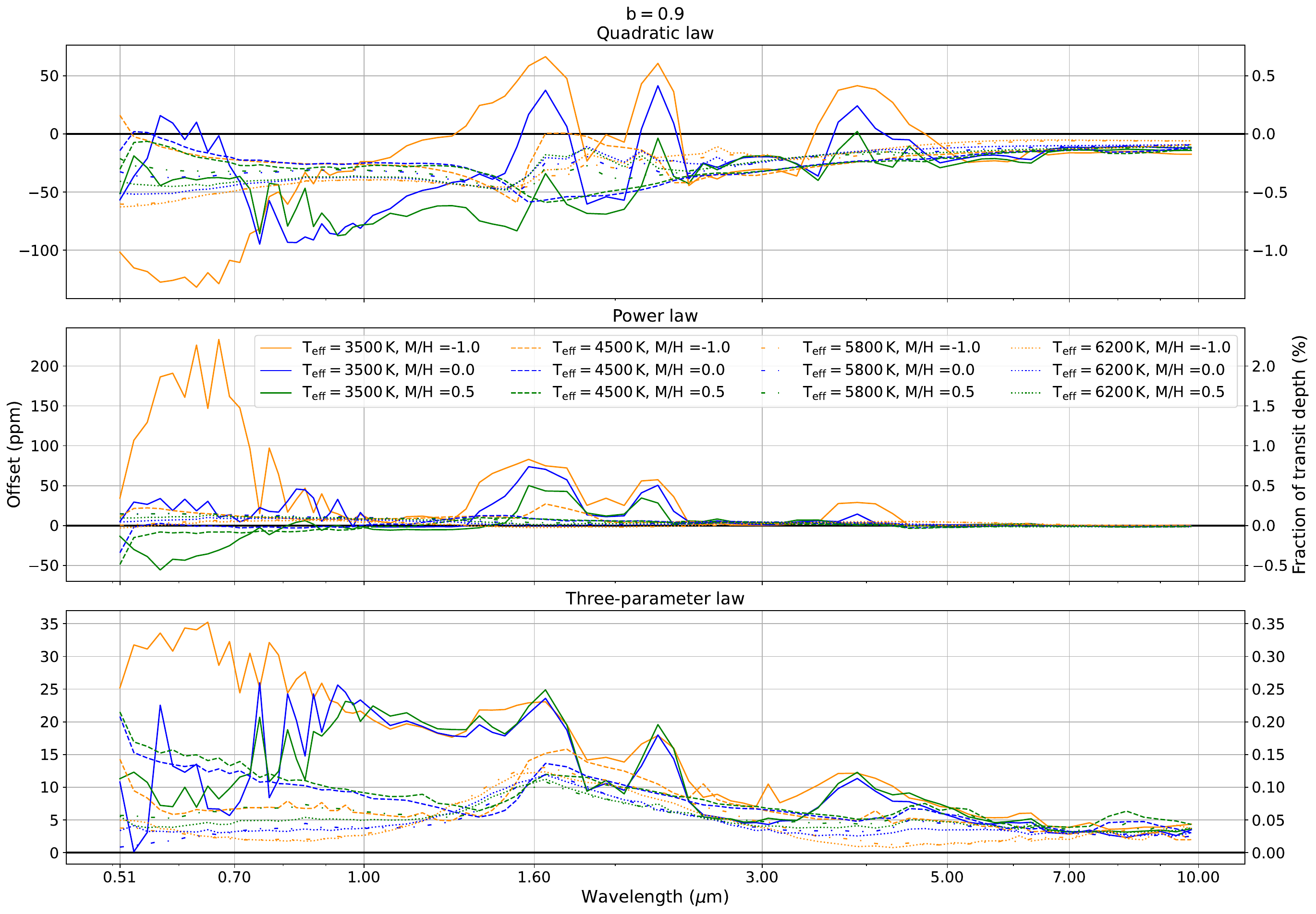}
    \caption{Offsets from true transit depth, for a planet transiting at $b = 0.9$, for all twelve stars.}
    \label{fig:Teffvar_mhvar_b0.9}
\end{sidewaysfigure}
\FloatBarrier

\FloatBarrier


\bibliography{main, research_doc}{}
\bibliographystyle{aa_url.bst}



\end{document}